\documentclass[12pt]{article}

\begin{document}

\begin{center}
{\Large\bf{}Black hole's tidal heating and angular momentum}\\
\end{center}

\begin{center}
Lau Loi So  
\end{center}

\begin{abstract}
In 1985 Thorne and Hartle used the Landau-Lifshitz pseudotensor to
demonstrate the tidal heating and angular momentum flux for a
black hole. Later in 2004, Poisson used the gravitational
perturbation method to study a black hole and obtained the same
result. Poisson proposed a new idea, that the mass quardupole
moment and current quadrupole moment can be written as the rate of
change of the tidal gravitational field. Inspired by these two
papers, we use the method of Thorne and Hartle to study other
classical pseudotensors: Einstein, Bergmann-Thomson, Papapetrou
and Weinberg. Moreover, we also constructed a general expression
pseudotensor. We find that for (i) tidal heating: other classical
pseudotensors give the same result as the Landau-Lifshitz
contribution. (ii) angular momentum flux: except for the Einstein
pseudotensor, all of them give the same value as the
Landau-Lifshitz pseudotensor.
\end{abstract}

\section{Introduction}

To examine the black hole tidal heating and angular momentum flux,
Thorne and Hartle used the Landau-Lifshitz pseudotensor~\cite{LL}
to calculate these two physical quantities in 1985~\cite{Thorne}.
Instead of using this classical pseudotensor, Poisson used the
gravitational perturbation method to study the tidal heating and
angular momentum flux of a black hole in 2004~\cite{Poisson}. It
is not surprising that Poisson obtained the same results as Thorne
did, but the surprising thing is that Poisson proposed a new
angle. In his paper: ``This is a rather surprising result, as one
would expect the quadrupole deformation of a tidally-distorted
body to be proportional to the tidal gravitational field itself,
instead of its time derivative." Because of this new perspective,
one can immediately find that the rate of change of the energy
that is absorbed by a black hole, from the outside universe, is
non-negative.

The classical pseudotensors include those of
Einstein~\cite{Einstein}, Landau-Lifshitz,
Bergmann-Thomson~\cite{BT}, Papapetrou~\cite{Papapetrou} and
Weinberg~\cite{Weinberg}. Moreover, we also constructed a general
expression pseudotensor. Inspired by the work of Thorne and
Hartle, our present paper will demonstrate the same physical
quantities, the tidal heating and angular momentum flux of a black
hole using other classical pseudotensors. We find that for (i)
tidal heating: other classical pseudotensors give the same result
as the Landau-Lifshitz, (ii) the angular momentum: except for the
Einstein pseudotensor, all of them contribute the same value as
the Landau-Lifshitz pseudotensor.

\section{Technical background}

Throughout this work, we use the same spacetime signature and
notation as in~\cite{MTW}, and we use the geometrical units
$G=c=1$, where $G$ and $c$ are the Newtonian constant and the
speed of light. The Greek letters denote the spacetime and Latin
letters refer to spatial. In principle, a classical
pseudotensor~\cite{CQGSoNesterChen2009} can be obtained from a
rearrangement of the Einstein field equation:
$G^{\mu\nu}=\kappa\,T^{\mu\nu}$, where the constant
$\kappa=8\pi{}G/c^{4}$, $G^{\mu\nu}$ and $T^{\mu\nu}$ are the
Einstein and stress tensors.  We define the gravitational
energy-momentum density pseudotensor in terms of a suitable
superpotential $U_{\alpha}{}^{[\mu\nu]}$:
\begin{eqnarray}
2\kappa\sqrt{-g}\,t_{\alpha}{}^{\mu}:=\partial_{\nu}U_{\alpha}{}^{[\mu\nu]}-2\sqrt{-g}\,G_{\alpha}{}^{\mu}.
\label{6aMar2015}
\end{eqnarray}
Alternatively, one can rewrite (\ref{6aMar2015}) as
$\partial_{\nu}U_{\alpha}{}^{[\mu\nu]}
=2\sqrt{-g}{}(G_{\alpha}{}^{\mu}+\kappa\,t_{\alpha}{}^{\mu})$.
Using the Einstein equation, the total energy-momentum density
complex can be defined as
\begin{eqnarray}
{\cal{T}}_{\alpha}{}^{\mu}:=\sqrt{-g}(T_{\alpha}{}^{\mu}+t_{\alpha}{}^{\mu})
=(2\kappa)^{-1}\partial_{\nu}U_{\alpha}{}^{[\mu\nu]}.
\end{eqnarray}
This total energy-momentum density is automatically conserved as
$\partial_{\mu}{\cal{T}}_{\alpha}{}^{\mu}\equiv{}0$, which can be
split into two parts:
\begin{eqnarray}
\partial_{\nu}U_{\alpha}{}^{[\mu\nu]}=2\sqrt{-g}\,G_{\alpha}{}^{\mu},\quad{}
\partial_{\nu}U_{\alpha}{}^{[\mu\nu]}=2\sqrt{-g}\kappa\,t_{\alpha}{}^{\mu}.\label{9bMar2015}
\end{eqnarray}
The first part indicates the contribution of matter and the second
piece refers to vacuum gravity~\cite{MTW}. For the condition of
interior mass-energy density, it is known that the Einstin,
Landau-Lifshitz, Bergmann-Thomson, Papapetrou and Weinberg
pseudotensors all satisfy the standard result
$2G_{\alpha}{}^{\mu}$ at the origin in Riemann normal coordinates.

Actually there are three kinds of superpotentials:
$U_{\alpha}{}^{[\mu\nu]}$, $U^{\alpha[\mu\nu]}$ and
$H^{[\alpha\beta][\mu\nu]}$. This also must satisfy some further
symmetries, under the interchange of the pairs, and the total
antisymmetry of 3 indices.  From them the total energy-momentum
complex can be obtained as follows:
\begin{eqnarray}
2\kappa{\cal{T}}_{\alpha}{}^{\mu}=\partial_{\nu}U_{\alpha}{}^{[\mu\nu]},\quad{}
2\kappa{\cal{T}}^{\alpha\mu}=\partial_{\nu}U^{\alpha[\mu\nu]},\quad{}
2\kappa{\cal{T}}^{\alpha\mu}=\partial^{2}_{\beta\nu}H^{[\alpha\beta][\mu\nu]}.
\end{eqnarray}
In vacuum, $t^{0j}$ is the gravitational energy flux and $t^{kf}$
is the angular momentum flux~\cite{Thorne}. The tidal heating and
angular momentum can be computed as
\begin{eqnarray}
\frac{dM}{dt}=-\frac{1}{2\kappa}\oint_{\partial{}V}\sqrt{-g}\,t^{0j}\,\hat{n}_{j}\,r^{2}\,d\Omega,\quad
\frac{dJ^{i}}{dt}=-\frac{1}{2\kappa}\oint_{\partial{}V}\epsilon^{i}{}_{j\,k}x^{j}
\sqrt{-g}\,t^{kf}\hat{n}_{f}\,r^{2}\,d\Omega,
\end{eqnarray}
where $r\equiv\sqrt{\delta_{ab}x^{a}x^{b}}$ is the distance from
the body in its local asymptotic rest frame and
$\hat{n}_{j}\equiv{}x_{j}/r$ is the unit radial vector.

In weak field the metric tensor can be decomposed as
$g_{\alpha\beta}=\eta_{\alpha\beta}+h_{\alpha\beta}$, and its
inverse $g^{\alpha\beta}=\eta^{\alpha\beta}-h^{\alpha\beta}$. Here
we mainly use the first order and ignore the higher orders.
According to Zhang~\cite{Zhang}, the metric components can be
written as
\begin{eqnarray}
h^{00}&=&\frac{3}{r^{5}}I_{ab}x^{a}x^{b}-E_{ab}x^{a}x^{b},\\
h^{0j}&=&\frac{4}{r^{5}}\,\epsilon^{j}{}_{pq}J^{p}{}_{l}\,x^{q}x^{l}
+\frac{2}{3}\epsilon^{j}{}_{pq}B^{p}{}_{l}x^{q}x^{l}
+\frac{2}{r^{3}}\dot{I}^{j}{}_{a}\,x^{a}
+\frac{10}{21}\dot{E}_{ab}x^{a}x^{b}x^{j}
-\frac{4}{21}\dot{E}^{j}{}_{a}x^{a}r^{2},\\
h^{ij}&=&\eta^{ij}h^{00}+\bar{h}^{ij},
\end{eqnarray}
where
\begin{eqnarray}
\bar{h}^{ij}=\frac{8}{3r^{3}}\epsilon_{pq}{}^{(i}\dot{J}^{j)p}x^{q}
+\frac{5}{21}x^{(i}\epsilon^{j)}{}_{pq}\dot{B}^{q}{}_{l}x^{p}x^{l}
-\frac{1}{21}r^{2}\epsilon_{pq}{}^{(i}\dot{B}^{j)\,q}x^{p}.
\end{eqnarray}
Zhang used $\bar{h}^{\alpha\beta}$ for the manipulation while we
prefer using $h^{\alpha\beta}$, the transformation is as follows
\begin{eqnarray}
\bar{h}^{\alpha\beta}=h^{\alpha\beta}-\frac{1}{2}\eta^{\alpha\beta}h.
\end{eqnarray}
The corresponding first order harmonic gauge is
$\partial_{\beta}\bar{h}^{\alpha\beta}=0$. Moreover, we will
substitute the mass quadrupole moment $I_{ij}$ and current
quadrupole moment $J_{ij}$ as determined by
Poisson~\cite{Poisson}:
\begin{eqnarray}
I^{ij}=\frac{32}{45}M^{6}\dot{E}^{ij},\quad{}
J^{ij}=\frac{8}{15}M^{6}\dot{B}^{ij},
\end{eqnarray}
where $M$ is the mass of the black hole.

\section{The classical pseudotensors}
Tidal heating and the angular momentum momentum flux for the
classical pseudotensors are presented.

\subsection{Einstein pseudotensor}
The following is the Freud (F) superpotential:
\begin{eqnarray}
_{F}U_{\alpha}{}^{[\mu\nu]}:=-\sqrt{-g}g^{\beta\sigma}\Gamma^{\tau}{}_{\lambda\beta}
\delta^{\lambda\mu\nu}_{\tau\sigma\alpha}.
\end{eqnarray}
The corresponding Einstein pseudotensor is
\begin{eqnarray}
_{E}t_{\alpha}{}^{\mu}
&=&2G_{\alpha}{}^{\mu}+\delta^{\mu}_{\alpha}(\Gamma^{\beta\lambda}{}_{\nu}\Gamma^{\nu}{}_{\beta\lambda}
-\Gamma^{\beta}{}_{\beta\nu}\Gamma^{\nu\lambda}{}_{\lambda})
+\Gamma^{\beta}{}_{\beta\alpha}\Gamma^{\mu\lambda}{}_{\lambda}\nonumber\\
&&+(\Gamma^{\mu\nu}{}_{\alpha}+\Gamma^{\nu\mu}{}_{\alpha})\Gamma^{\beta}{}_{\beta\nu}
-\Gamma^{\beta}{}_{\beta\alpha}\Gamma^{\lambda\mu}{}_{\lambda}
-2\Gamma^{\beta\nu}{}_{\alpha}\Gamma^{\mu}{}_{\beta\nu}.
\end{eqnarray}
Using this in~\cite{BT} in vacuum, the tidal heating for the black
hole is
\begin{eqnarray}
\frac{dM}{dt}&=&\frac{1}{2}\left(I^{ij}\dot{E}_{ij}+\frac{4}{3}J^{ij}\dot{B}_{ij}\right)\nonumber\\
&=&\frac{16}{45}M^{6}(\dot{E}^{ij}\dot{E}_{ij}+\dot{H}^{ij}\dot{H}_{ij}),
\end{eqnarray}
and the angular momentum flux is
\begin{eqnarray}
\frac{dJ^{i}}{dt}&=&-\epsilon^{i}{}_{ab}\left(I^{ac}E^{b}{}_{c}+\frac{10}{9}J^{ac}B^{b}{}_{c}\right)\nonumber\\
&=&-\frac{32}{45}M^{6}\epsilon^{i}{}_{ab}\left(\dot{E}^{ac}E^{b}{}_{c}+\frac{5}{6}\dot{B}^{ac}B^{b}{}_{c}\right).
\end{eqnarray}
This result for angular momentum flux is not the same as that for
the other classical pseudotensors, as we shall see below.

\subsection{Landau-Lifshitz pseudotensor}
Here we repeat the calculation as Thorne and Hartle did before,
but we have different algebra since we used our notation. Recall
the Landau-Lifshitz (LL) superpotential
\begin{eqnarray}
U^{\alpha[\mu\nu]}_{LL}:=-(-g)g^{\alpha\beta}g^{\pi\sigma}
\Gamma^{\tau}{}_{\lambda\pi}\delta^{\lambda\mu\nu}_{\tau\sigma\beta},
\end{eqnarray}
and the pseudotensor becomes
\begin{eqnarray}
t^{\alpha\mu}_{LL}&=&2\,G^{\alpha\mu}
+g^{\alpha\mu}\,\Gamma^{\gamma}{}_{\gamma\nu}\Gamma^{\pi}{}_{\pi}{}^{\nu}
-2g^{\alpha\mu}\,\Gamma^{\gamma}{}_{\gamma\nu}\Gamma^{\nu\pi}{}_{\pi}
+g^{\alpha\mu}\Gamma^{\nu\,\rho\,\pi}\Gamma_{\rho\,\nu\,\pi}\nonumber\\
&&+\Gamma^{\alpha\mu}{}_{\nu}\,\Gamma^{\nu\pi}{}_{\pi}
-\Gamma^{\alpha\mu\rho}\Gamma^{\pi}{}_{\pi\rho}
-\Gamma^{\mu\alpha}{}_{\nu}\Gamma^{\gamma}{}_{\gamma}{}^{\nu}
+\Gamma^{\mu\alpha}{}_{\rho}\,\Gamma^{\rho\,\nu}{}_{\nu}
+2\Gamma^{\nu\alpha\mu}\Gamma^{\gamma}{}_{\gamma\,\nu}\nonumber\\
&&+\Gamma^{\alpha\pi\rho}\Gamma^{\mu}{}_{\pi\rho}
-\Gamma^{\alpha\nu}{}_{\nu}\,\Gamma^{\mu\pi}{}_{\pi}
+\Gamma^{\alpha\nu}{}_{\nu}\,\Gamma^{\pi}{}_{\pi}{}^{\mu}
-\Gamma^{\alpha\nu\pi}\Gamma_{\nu\pi}{}^{\mu}\nonumber\\
&&+\Gamma^{\gamma}{}_{\gamma}{}^{\alpha}\,\Gamma^{\mu\pi}{}_{\pi}
-\Gamma^{\rho\pi\alpha}\Gamma^{\mu}{}_{\rho\pi}
-\Gamma^{\pi\nu\,\alpha}\Gamma_{\nu\,\pi}{}^{\mu}
-\Gamma^{\gamma}{}_{\gamma}{}^{\alpha}\Gamma^{\pi}{}_{\pi}{}^{\mu},
\end{eqnarray}
which is symmetric in $\alpha$ and $\mu$. For this pseudotensor
the tidal heating works out to be
\begin{eqnarray}
\frac{dM}{dt}&=&\frac{1}{2}\left(I^{ij}\dot{E}_{ij}+\frac{4}{3}J^{ij}\dot{B}_{ij}\right)\nonumber\\
&=&\frac{16}{45}M^{6}(\dot{E}^{ij}\dot{E}_{ij}+\dot{H}^{ij}\dot{H}_{ij}),
\end{eqnarray}
and the angular momentum flux is
\begin{eqnarray}
\frac{dJ^{i}}{dt}&=&-\epsilon^{i}{}_{ab}\left(I^{ac}E^{b}{}_{c}+\frac{4}{3}J^{ac}B^{b}{}_{c}\right)\nonumber\\
&=&-\frac{32}{45}M^{6}\epsilon^{i}{}_{ab}(\dot{E}^{ac}E^{b}{}_{c}+\dot{B}^{ac}B^{b}{}_{c}).
\end{eqnarray}

\subsection{Bergmann-Thomson pseudotensor}
The Bergmann-Thomson (BT) superpotential is defined as
\begin{eqnarray}
U^{\alpha[\mu\nu]}_{BT}:=-\sqrt{-g}g^{\alpha\beta}g^{\pi\sigma}
\Gamma^{\tau}{}_{\lambda\pi}\delta^{\lambda\mu\nu}_{\tau\sigma\beta},
\end{eqnarray}
and the associated pseudotensor is
\begin{eqnarray}
t^{\alpha\mu}_{BT}
&=&2G^{\alpha\mu}-(g^{\alpha\mu}\Gamma^{\pi}{}_{\pi\rho}
-\Gamma^{\alpha\mu}{}_{\rho}-\Gamma^{\mu\alpha}{}_{\rho})\Gamma^{\rho\nu}{}_{\nu}
+\Gamma^{\alpha\nu}{}_{\nu}(\Gamma^{\pi\mu}{}_{\pi}-\Gamma^{\mu\pi}{}_{\pi})
+g^{\alpha\mu}\Gamma^{\beta\nu}{}_{\rho}\Gamma^{\rho}{}_{\beta\nu}\nonumber\\
&&+(\Gamma^{\rho\alpha\mu}-\Gamma^{\alpha\mu\rho})\Gamma^{\pi}{}_{\pi\rho}
+\Gamma^{\alpha\pi\rho}(\Gamma^{\mu}{}_{\pi\rho}-\Gamma_{\pi\rho}{}^{\mu})
-\Gamma^{\rho\pi\alpha}(\Gamma^{\mu}{}_{\rho\pi}+\Gamma_{\pi\rho}{}^{\mu}).
\end{eqnarray}
In vacuum, using this expression the tidal heating works out to be
\begin{eqnarray}
\frac{dM}{dt}&=&\frac{1}{2}\left(I^{ij}\dot{E}_{ij}+\frac{4}{3}J^{ij}\dot{B}_{ij}\right)\nonumber\\
&=&\frac{16}{45}M^{6}(\dot{E}^{ij}\dot{E}_{ij}+\dot{H}^{ij}\dot{H}_{ij}),
\end{eqnarray}
and the angular momentum flux is
\begin{eqnarray}
\frac{dJ^{i}}{dt}&=&-\epsilon^{i}{}_{ab}\left(I^{ac}E^{b}{}_{c}+\frac{4}{3}J^{ac}B^{b}{}_{c}\right)\nonumber\\
&=&-\frac{32}{45}M^{6}\epsilon^{i}{}_{ab}(\dot{E}^{ac}E^{b}{}_{c}+\dot{B}^{ac}B^{b}{}_{c}).
\end{eqnarray}

\subsection{Papapetrou pseudotensor}
The Papapertrou superpotential is defined as
$H^{[\mu\nu][\alpha\beta]}_{P}
:=-\sqrt{-g}g^{\rho\pi}\eta^{\tau\gamma}\delta_{\pi\gamma}^{\nu\mu}\delta_{\rho\tau}^{\alpha\beta}$,
equivalently one can use
$U^{\alpha[\mu\nu]}_{P}\equiv\partial_{\beta}H^{[\mu\nu][\alpha\beta]}$.
In terms of the Bergmann-Thomson superpotential
$U_{BT}^{\alpha[\mu\nu]}$, we have
\begin{eqnarray}
U_{P}^{\alpha[\mu\nu]}:=U_{BT}^{\alpha[\mu\nu]}+\sqrt{-g}\sum^{4}_{i=1}U_{i},
\end{eqnarray}
where $U_{1}$ to $U_{4}$ are given below in Table 1. In vacuum,
the tidal heating using this pesudotensor (i.e.,
$t^{\alpha\mu}_{P}=\partial_{\nu}U^{\alpha[\mu\nu]}_{P}$) works
out to be
\begin{eqnarray}
\frac{dM}{dt}&=&\frac{1}{2}\left(I^{ij}\dot{E}_{ij}+\frac{4}{3}J^{ij}\dot{B}_{ij}\right)\nonumber\\
&=&\frac{16}{45}M^{6}(\dot{E}^{ij}\dot{E}_{ij}+\dot{H}^{ij}\dot{H}_{ij}),
\end{eqnarray}
and the angular momentum flux is
\begin{eqnarray}
\frac{dJ^{i}}{dt}&=&-\epsilon^{i}{}_{ab}\left(I^{ac}E^{b}{}_{c}+\frac{4}{3}J^{ac}B^{b}{}_{c}\right)\nonumber\\
&=&-\frac{32}{45}M^{6}\epsilon^{i}{}_{ab}(\dot{E}^{ac}E^{b}{}_{c}+\dot{B}^{ac}B^{b}{}_{c}).
\end{eqnarray}

\subsection{Weinberg pseudotensor}
The superpotential for Weinberg(W) is
\begin{eqnarray}
H^{[\mu\nu][\alpha\beta]}_{W}
:=\sqrt{-\eta}\,\eta^{\alpha\pi}\eta^{\beta\xi}\eta^{\lambda\kappa}\delta^{\sigma\mu\nu}_{\pi\xi\kappa}g_{\lambda\sigma}.
\end{eqnarray}
An alternative representation for this superpotential is
\begin{eqnarray}
U^{\alpha[\mu\nu]}_{W}:=U^{\alpha[\mu\nu]}_{W_{0}}+\sqrt{-\eta}\sum^{10}_{i=2}U_{i},\label{27aDec2019}
\end{eqnarray}
where the detailed expressions from $U_{2}$ to $U_{10}$ are given
below in Table 1, and
\begin{eqnarray}
U^{\alpha[\mu\nu]}_{W_{0}}
:=-\sqrt{-\eta}\,g^{\alpha\beta}g^{\pi\sigma}
\Gamma^{\tau}{}_{\lambda\pi}\delta^{\lambda\mu\nu}_{\tau\sigma\beta},
\end{eqnarray}
and its associated pseudotensor is
\begin{eqnarray}
t^{\alpha\mu}_{W_{0}}&=&2G^{\alpha\mu}
+g^{\alpha\mu}(\Gamma^{\beta\lambda}{}_{\nu}\Gamma^{\nu}{}_{\beta\lambda}
-\Gamma^{\nu\lambda}{}_{\nu}\Gamma^{\beta}{}_{\beta\lambda})\nonumber\\
&&-\Gamma^{\alpha\mu}{}_{\nu}\Gamma^{\beta\nu}{}_{\beta}
+\Gamma^{\alpha\mu}{}_{\nu}\Gamma^{\nu\lambda}{}_{\lambda}
+\Gamma^{\mu\alpha\lambda}\Gamma^{\beta}{}_{\beta\lambda}
+\Gamma^{\mu\alpha}{}_{\lambda}\Gamma^{\lambda\nu}{}_{\nu}
\nonumber\\
&&-\Gamma^{\alpha\nu}{}_{\nu}\Gamma^{\mu\lambda}{}_{\lambda}
+\Gamma^{\alpha\beta\nu}\Gamma^{\mu}{}_{\beta\nu}
+\Gamma^{\alpha\nu}{}_{\nu}\Gamma^{\beta\mu}{}_{\beta}
-\Gamma^{\alpha\beta\nu}\Gamma_{\beta\nu}{}^{\mu}
\nonumber\\
&&-\Gamma^{\beta\nu\alpha}\Gamma^{\mu}{}_{\beta\nu}
-\Gamma^{\nu\alpha}{}_{\nu}\Gamma^{\mu\lambda}{}_{\lambda}
+\Gamma^{\nu\alpha}{}_{\nu}\Gamma^{\beta\mu}{}_{\beta}
-\Gamma^{\beta\nu\alpha}\Gamma_{\nu\beta}{}^{\mu}.
\end{eqnarray}
Referring to (\ref{27aDec2019}), in vacuum, the tidal heating is
\begin{eqnarray}
\frac{dM}{dt}&=&\frac{1}{2}\left(I^{ij}\dot{E}_{ij}+\frac{4}{3}J^{ij}\dot{B}_{ij}\right)\nonumber\\
&=&\frac{16}{45}M^{6}(\dot{E}^{ij}\dot{E}_{ij}+\dot{H}^{ij}\dot{H}_{ij}),
\end{eqnarray}
and the angular momentum flux is
\begin{eqnarray}
\frac{dJ^{i}}{dt}&=&-\epsilon^{i}{}_{ab}\left(I^{ac}E^{b}{}_{c}+\frac{4}{3}J^{ac}B^{b}{}_{c}\right)\nonumber\\
&=&-\frac{32}{45}M^{6}\epsilon^{i}{}_{ab}(\dot{E}^{ac}E^{b}{}_{c}+\dot{B}^{ac}B^{b}{}_{c}).
\end{eqnarray}

\subsection{General expression pseudotensor}
For simplicity, we use the Bergmann-Thomson superpotential as a
leading term and consider the 13 linear artificial higher order
term combinations as a modification:
\begin{eqnarray}
U^{\alpha[\mu\nu]}:=U^{\alpha[\mu\nu]}_{BT}+\sqrt{-g}\sum_{i=1}^{13}c_{i}U_{i},\label{23Mar2016}
\end{eqnarray}
where the $c_{i}$ are finite constants, and the explicit extra
terms are shown in Table 1. Note that both
$\partial_{0}(I^{ij}E_{ij})$ and $\partial_{0}(J^{ij}B_{ij})$
contribute null result for the tidal work dissipation, where
$\partial_{0}$ means the time derivative. In (\ref{23Mar2016}),
consider within a region with a non-vanishing matter tensor, then
the leading of the Bergmann-Thomson superpotential gives the
standard value $2G_{\alpha}{}^{\mu}$, but all the $U_{i}$
contribute null result to the lowest order. In vacuum, the results
of the extra tidal heating work and angular momentum flux are
shown in Table 1. The tidal heating is the same as the
Bergmann-Thomson, i.e.,
\begin{eqnarray}
\frac{dM}{dt}=\frac{16}{45}M^{6}(\dot{E}^{ij}\dot{E}_{ij}+\dot{H}^{ij}\dot{H}_{ij}).
\end{eqnarray}
But the angular momentum flux is
\begin{eqnarray}
\frac{dJ^{i}}{dt}=-\frac{32}{45}M^{6}\epsilon^{i}{}_{ab}
\left[\dot{E}^{ac}E^{b}{}_{c}+\left(1+\frac{\alpha}{12}\right)\dot{B}^{ac}B^{b}{}_{c}\right],
\end{eqnarray}
where $\alpha=-c_{1}+c_{2}-2c_{9}+c_{10}$. This means that the
tidal heating work is completely determined by the leading
Bergmann-Thomson pseudotensor. However, for the angular momentum
flux, we can alter the value by tuning these four arbitrary
constants $c_{1},c_{2}, c_{9}$ and $c_{10}$. Perhaps, the simplest
case may be when $\alpha=0$. Thus, for this general pseudotensor
expression, the result for the tidal heating and angular momentum
flux are the same as those given by the Bergmann-Thomson
expression, i.e., the standard values.

\begin{table}[ht]
  \caption{Tidal heating work and angular momentum flux for $\sqrt{-g}\,U_{i}$}
\centering
\begin{tabular}{ccc}
\hline\hline Superpotential & Tidal heating work & Angular
momentum flux \\ [0.5ex] \hline $U_{1}=
h^{\mu\pi}\Gamma^{\alpha\nu}{}_{\pi}-(\mu\leftrightarrow\nu)$ &
$-\frac{1}{10}\partial_{0}(I^{ij}E_{ij})$ & $\frac{1}{9}\epsilon^{i}{}_{ab}J^{ac}B^{b}{}_{c}$ \\
$U_{2}=h^{\mu\pi}\Gamma^{\nu\alpha}{}_{\pi}-(\mu\leftrightarrow\nu)
$ & $-\frac{1}{15}\partial_{0}(J^{ij}B_{ij})$  & $-\frac{1}{9}\epsilon^{i}{}_{ab}J^{ac}B^{b}{}_{c}$ \\
$U_{3}=g^{\alpha\mu}h^{\beta\nu}\Gamma^{\lambda}{}_{\lambda\beta}-(\mu\leftrightarrow\nu)
$ & $-\frac{1}{10}\partial_{0}(I^{ij}E_{ij})$  & 0 \\
$U_{4}=h^{\alpha\nu}\Gamma^{\mu\lambda}{}_{\lambda}-(\mu\leftrightarrow\nu)
$ & 0  & 0 \\
$U_{5}=h^{\alpha\mu}\Gamma^{\lambda\nu}{}_{\lambda}-(\mu\leftrightarrow\nu)
$ & $\frac{1}{10}\partial_{0}(I^{ij}E_{ij})$  & 0 \\
$U_{6}=g^{\alpha\mu}h^{\beta\lambda}\Gamma_{\beta\lambda}{}^{\nu}-(\mu\leftrightarrow\nu)
$ & $-\frac{1}{5}\partial_{0}(I^{ij}E_{ij}+\frac{2}{3}J^{ij}B_{ij})$  & 0 \\
$U_{7}=g^{\alpha\nu}h^{\beta\mu}\Gamma_{\beta}{}^{\lambda}{}_{\lambda}-(\mu\leftrightarrow\nu)
$ & 0  & 0 \\
$U_{8}=g^{\alpha\nu}h^{\beta\lambda}\Gamma^{\mu}{}_{\beta\lambda}-(\mu\leftrightarrow\nu)
$ &
$-\frac{1}{10}\partial_{0}(I^{ij}E_{ij}-\frac{4}{3}J^{ij}B_{ij})$
&  0\\
$U_{9}=h^{\alpha\lambda}\Gamma^{\nu\mu}{}_{\lambda}-(\mu\leftrightarrow\nu)
$ &
$\frac{1}{10}\partial_{0}(I^{ij}E_{ij}+\frac{4}{3}J^{ij}B_{ij})$
& $\frac{2}{9}\epsilon^{i}{}_{ab}J^{ac}B^{b}{}_{c}$ \\
$U_{10}=h^{\nu\pi}\Gamma_{\pi}{}^{\alpha\mu}-(\mu\leftrightarrow\nu)
$ & $-\frac{1}{15}\partial_{0}(J^{ij}B_{ij})$ &
$-\frac{1}{9}\epsilon^{i}{}_{ab}J^{ac}B^{b}{}_{c}$ \\
$U_{11}=h\Gamma^{\nu\mu\alpha}-(\mu\leftrightarrow\nu) $ &
$-\frac{1}{5}\partial_{0}(I^{ij}E_{ij})$   & 0 \\
$U_{12}=g^{\alpha\mu}h\Gamma^{\lambda\nu}{}_{\lambda}-(\mu\leftrightarrow\nu)
$ & $-\frac{1}{5}\partial_{0}(I^{ij}E_{ij})$   & 0 \\
$U_{13}=g^{\alpha\mu}h\Gamma^{\nu\lambda}{}_{\lambda}-(\mu\leftrightarrow\nu)
$ & 0  & 0 \\ [1ex] \hline
\end{tabular}
\end{table}

\section{Conclusion}
In our universe, black holes are some of the strangest and most
fascinating objects. In 1985 Thorne and Hartle used the
Landau-Lifshitz pseudotensor to study the tidal heating and
angular momentum flux for a black hole. After 19 years, Poisson
used an alternative method, the gravitational perturbation, to
study the black hole and obtained the same result. Our present
paper use the method as Thorne and Hartle did, demonstrate the
Einstein, Bergmann-Thomson, Papapetrou and Weinberg pseudotensors.
Moreover, we also constructed a general expression pseudotensor.
We find that, for the tidal heating work, all of these
pseudotensors give the same value as Landau-Lifshitz. However, for
the angular momentum flux, except Einstein pseudotensor gives a
slightly different value, and all the other classical
pseudotensors contribute the same result. Moreover, we also
constructed a general expression pseudotensor to illustrate the
other possible combinations.

\end{document}